# Anti-crossing properties of strong coupling system of silver nanoparticle dimers coated with thin dye molecular films analyzed by classical electromagnetism

Tamitake Itoh[1]*, Yuko S. Yamamoto[2], Takayuki Okamoto[3]

[1]Nano-Bioanalysis Research Group, Health Research Institute, National Institute of Advanced Industrial Science and Technology (AIST), Takamatsu, Kagawa 761-0395, Japan

[2]School of Materials Science, Japan Advanced Institute of Science and Technology (JAIST), Nomi, Ishikawa 923-1292, Japan

[3]Advanced Device Laboratory, RIKEN, Hirosawa, Wako, Saitama 351-0198, Japan

*Corresponding author: tamitake-itou@aist.go.jp

## ABSTRACT

The evidence of strong coupling between plasmons and molecular excitons for plasmonic nanoparticle (NP) dimers exhibiting ultra-sensitive surface enhanced resonant Raman scattering is the observation of anti-crossing in the coupled resonance.

However, it is not easy to experimentally tune plasmon resonance of such dimers for the observation. In this work, we theoretically investigate the anti-crossing properties of the dimers coated by the thin dye films with thicknesses greater than 0.1 nm and gap distances larger than 1.2 nm according to the principles of classical electromagnetism. The plasmon resonance spectra of these dimers are strongly affected by their coupling with the exciton resonance of dye molecules. A comparison of the film thickness dependences of dimer spectral changes with those of silver ellipsoidal NPs indicates that the dipole plasmons localized in the dimer gap are coupled with molecular excitons of the film much stronger than the dipole plasmons of ellipsoidal NPs. Furthermore, the anti-crossing of coupled resonances is investigated while tuning plasmon resonance by changing the morphology and refractive index of the surrounding medium. The spectral changes observed for ellipsoidal NPs clearly exhibit anti-crossing properties; however, the anti-crossing behavior of dimers is more complex due to the strong coupling of dipoles and higher order plasmons with multiple molecular excitons. We find the anti-crossing for dimers is clearly confirmed by the refractive index dependence of coupled resonance.

## I. INTRODUCTION

Surface-enhanced resonant Raman scattering (SERRS) of single plasmonic nanoparticle (NP) systems enables us to directly observe relationships between plasmon resonance and SERRS, because we can identify "resonance" inducing "enhancement".[1–3] The relationships have been experimentally observed as unique features in the corresponding Raman spectral changes induced by plasmon resonance and quantitatively clarified based on the electromagnetic (EM) coupling between plasmons and dye molecular excitons.[4–6] During the investigations, plasmon resonance was found to exhibit blue shifts due to decreased SERRS activity, as shown in Fig. 1(a1).[7] These blue shifts are clarified as the decrease in the EM coupling energy using a coupled oscillator model composed of plasmonic and molecular exciton oscillators.[8–10] The resulting coupling energy reaches a magnitude of several hundred milli-electron volts, as shown in Fig. 1(a2).[8] Under near-single molecule (SM) SERRS conditions corresponding to low molecular concentrations ($< 10^{-9}$ M),[11,12] the obtained values of coupling energy indicate that a molecular dipole interacts with a vacuum EM field confined by a plasmon dipole within a volume of several cubic nanometers inside the dimer gap called a "hotspot" (see Fig. 1(b)).[9] They also mean that the exchange rate of photons between plasmons and excitons is larger than their dephasing rates.[13] Such a fast coupling process occurring within the coherent times of both resonances is called

"strong coupling", during which plasmon resonance couples with the molecular exciton resonance.[13] Such coupled resonances at SERRS hotspots have been experimentally examined by several research groups,[14,15] and their optical properties significantly differ from those of the original plasmon and molecular exciton resonances.[16,17] Thus, SERRS hotspots serve as an important platform for applying the principles of cavity quantum electrodynamics (QED) to the recently discovered exotic photochemical and photophysical phenomena.[16,17]

The strong coupling at SERRS hotspots should be verified to eliminate other possible reasons for the occurrence of blue shifts in plasmon resonance (for example, the photo-thermal melting of dimers via SERRS quenching). This task can be performed by observing the anti-crossing behavior of coupled resonance.[13] Anti-crossing is a phenomenon demonstrating that two resonators (in this work, a plasmon resonator and a molecular exciton resonator) cannot have equal resonance energies even after tuning.[13] In the case of the coupling process induced by the fluctuations of a vacuum EM field, the coupling energy is proportional to the energy gap between the two resonances, as shown in Fig. 1(c).[18] Anti-crossing is manifested in the form of a dispersion relationship between the coupled and plasmon resonances observed when tuning the plasmon resonance energy against the exciton resonance energy (see Fig. 1(d)).[9] In the case of dimers, it is not easy to experimentally detect anti-crossing behavior by tuning plasmon resonance through the adjustment of various experimental parameters including NP sizes, shapes, and gap distances because the latter significantly change the coupling energy. Furthermore, the multi-level properties of molecules make anti-crossing characteristics very unclear.[9] Thus, it is important to theoretically examine the anti-crossing process using these parameters combined with the realistic exciton properties of molecules.

In this study, we theoretically investigated the anti-crossing properties of dye-coated dimers and ellipsoidal NPs by analyzing their elastic scattering spectra while changing three parameters: the dye film thicknesses, gap distances (the aspect ratios for ellipsoidal NPs), and refractive indexes of their surrounding media. In classical electromagnetism, the effect of a vacuum EM field is intrinsically included in the complex refractive index and structures of the studied materials of the system.[16–18] Indeed, the imaginary parts of the dielectric responses of the system in classical electromagnetism correspond to the de-excitation by vacuum EM field fluctuations in cavity QED.[13] In addition, we used an experimentally obtained complex refractive index of the dye film to accurately describe its multi-level properties.[6] By analyzing the spectral changes observed in coupled resonance during changing these parameters, it was found that the plasmon dipole localized in the dimer gap was coupled with a molecular exciton much stronger than the plasmon dipole of ellipsoidal NPs. The higher order plasmons of dimers exhibit more complex anti-crossing properties as compared with those of ellipsoidal NPs due to the multiple coupling of the dipole plasmon, higher order plasmons, and molecular excitons. Furthermore, the anti-crossing characteristics of the dye-coated dimers are most effectively evaluated by changing refractive indexes of their surrounding media.

## II. MODEL

The elastic light scattering of silver prolate ellipsoidal NPs and spherical NP dimers adsorbed by rhodamine 6G (R6G) dye molecules are examined theoretically. Silver NPs adsorbed by the dye molecules are represented as the NPs covered by the thin dye films with thickness $t_f$ assuming that these molecules uniformly cover the NP surface, as shown in Fig. 2(a1).[7,19] Dye molecules sparsely adsorbed on the NP surfaces are represented as a film thinner than the size of a single dye molecule (~1 nm). Figures 2(a2) and 2(a3) display the ellipsoidal NP and dimer covered with a dye film. The radius $r$ of the spherical NP is set to 30 nm, and the gap distance $d$ is maintained larger than 1.2 nm to avoid the formation of charge transfer plasmons that are capable of the coherent tunneling of surface conductive electrons by excitation light.[20] The volume of ellipsoidal NPs is set to that of the dimer calculated by the formula $\frac{4\pi}{3}a^2b = 2\frac{4\pi}{3}r^3 (= \text{const.})$ in order to quantitatively compare the effect of the dye exciton resonance of the coated film on the plasmon resonance of the dimers with the effect observed for ellipsoidal NPs. We also use the complex refractive index of silver $N_{Ag}$ reported in Ref. 21 (see Fig. 2(b1)). The complex refractive index of R6G $N_{R6G}$ is experimentally obtained by analyzing the refraction spectrum of a thick R6G film using the Kramers−Kronig relation (Fig. 2(b2)).[7] Note that the spectral shape of the imaginary part of $N_{R6G}$ (Im[$N_{R6G}$]) is not Lorentzian, but appears to be a superposition of several Lorentzian spectra representing the multi-level system according to the Franck–Condon mechanism.[9,22] Furthermore, Im[$N_{R6G}$] exhibits another broad peak at around 3.7 eV, indicating that the spectral changes observed in the coupled resonances are more complex than the changes detected for two-level systems. The polarizability $\alpha_{ell}$ of an ellipsoidal NP is calculated using the quasi-static approximation[23] because the size of ellipsoidal NPs is assumed to be much smaller than the excitation wavelength. However, in the case of dimers, EM fields are too tightly confined inside the gap indicating that the quasi-static approximation is not applicable. Thus, higher order plasmons should be considered when studying the dimer polarizability. The polarizability $\alpha_{dim}$ of an NP dimer is calculated in accordance with the extended Mie theory, as described in Ref. 24. During computations, the order of multipoles is determined by increasing its magnitude until the calculated polarizability becomes independent of the order. The shorter gap requires a larger order; for example, the order of 80 corresponds to $d = 1.2$ nm. Figures 2(c1) and 2(c2) show the spectra of the scattering cross sections $\sigma_{sca}(\omega) = (k^4/6\pi)|\alpha|^2$ of the ellipsoidal NP with an aspect ratio of 0.42 and dimer with $d = 1.2$ nm, where $\alpha$ is the polarizability. The spectrum of the ellipsoidal NP clearly shows a dipole plasmon resonance peak centered at 2.5 eV. However, the dimer contains both dipoles and higher order plasmons in the gap, suggesting that the coated dimers exhibit complex anti-crossing properties due to the multiple EM coupling of the dipole plasmon, multipole plasmons, and molecular excitons. Their investigation involves tuning plasmon resonance spectra across the exciton resonance spectrum of the dye film, which is performed by two different methods in this study. The first method involves varying the morphology of ellipsoidal NPs and the dimers, while the second method is based on changing the refractive index $N_m$ of the surrounding medium.

Plasmon resonance tuning is conducted at different aspect ratios of ellipsoidal NPs. Under the quasi-static approximation, the polarizability of uncoated ellipsoidal NPs is expressed as

$$\alpha_{ell} = \frac{4\pi}{3} a^2 b \frac{\varepsilon - \varepsilon_m}{\varepsilon_m + L(\varepsilon - \varepsilon_m)}, \quad (1)$$

where $L$ is the depolarization factor, $\varepsilon \left(= \varepsilon_1 + i\varepsilon_2 = N_{Ag}^2\right)$ is the relative permittivity of NPs, and $\varepsilon_m \left(= N_m^2 = 1.3^2\right)$ is the relative permittivity of the surrounding medium.[23] The value of $L$ is proportional to the ability of NPs to cancel the external field, indicating that $L \propto a/b$.[23] Under the resonance condition $\varepsilon_m + L(\varepsilon_1 - \varepsilon_m) = 0$, the resonance energy increases with increasing aspect ratio because the value of $\varepsilon_1 (= \text{Re}[N_{Ag}]^2 - \text{Im}[N_{Ag}]^2)$ decreases with photon energy (see Fig. 2(b1)). Figure 2(d1) shows the blue shifts of $\sigma_{sca}(\omega)$ observed with increasing aspect ratio $a/b$ from 0.13 to 1.0. In Fig. 2(d1), the plasmon resonance clearly crosses the Im[$N_{R6G}$] spectrum, which represents the exciton resonance of the R6G film, suggesting that its anti-crossing properties can be examined. The volume of ellipsoidal NPs is maintained constant in order to retain the total amount of dye molecules while changing the aspect ratio.

For the dimers, plasmon resonance tuning is performed at various gap distances. In this work, we focus only on dipole plasmons while excluding higher order plasmons to simplify the explanation procedure (in the actual calculation procedure, that latter are taken into account). The polarizability of the dipole plasmon of an uncoated dimer under the quasi-static approximation is defined as

$$\alpha_{dim} = \frac{8}{3}\pi r^3 \frac{\varepsilon - \varepsilon_m}{(1-2K)\varepsilon + (2-2K)\varepsilon_m}, \quad (2)$$

where $K = \left(\frac{r}{2r+d}\right)^3 < \frac{1}{2}$.[25] The value of $(2-2K)/(1-2K)$ increases with $K$; hence, the resonance energy satisfying the condition $(1-2K)\varepsilon + (2-2K)\varepsilon_m = 0$ increases with increasing $d$ because $\varepsilon_1 (= \text{Re}[N_{Ag}]^2 - \text{Im}[N_{Ag}]^2)$ is inversely proportional to the photon energy, as shown in Fig. 2(b1). The dipole plasmon resonance energy of dimers is determined by the dipole-dipole interaction energy, which is proportional to the third power of the reciprocal gap distance. Figure 2(d2) shows the blue shifts of $\sigma_{sca}(\omega)$ in the range of $d$ from 1.2 to 12.0 nm. Here, the plasmon resonance does not apparently cross the exciton resonance; however, the anti-crossing properties of the dimers may be observable.

For plasmon resonance tuning, the refractive indexes $N_m \left(= \sqrt{\varepsilon_m}\right)$ of the media surrounding ellipsoidal NPs and the dimers are changed as well. Several research groups have theoretically examined a tuning method that involved the observation of the anti-crossing phenomenon using monomer NPs and nano-cube dimers adsorbed by quantum dots and J-aggregates.[26,27] Equation (1) indicates that the resonance energy blueshifts with decreasing $N_m$ because the value of $\varepsilon_1$ should satisfy the resonance condition $\varepsilon_m + L(\varepsilon_1 - \varepsilon_m) = 0$. Similarly, in the case of dimers, Eq. (2) indicates that the resonance energy blueshifts with decreasing $N_m$ to satisfy the resonance condition $(1-2K)\varepsilon + (2-2K)\varepsilon_m = 0$. Figures 2(e1) and 2(e2) show that the higher energy shifts observed in the plasmon resonances of ellipsoidal NPs and dimers by decreasing $N_m$ from 2.0 to 1.0. The dipolar plasmon resonances of both structures clearly cross the exciton resonance of the dye films; thus, their anti-crossing properties can be examined at various values of $N_m$.

## III. RESULTS AND DISCUSSION

First, the sensitivity of the dipole plasmon resonance to the exciton resonance of the coated dye films was examined for ellipsoidal NPs and the dimers as a dependence of the plasmon resonance on the film thickness. Figure 3(a1) shows the changes in the $\sigma_{sca}(\omega)$ spectra of coated ellipsoidal NPs observed at $t_f$ values ranging from 0.0 to 10.0 nm. The aspect ratio is set to 0.42 to ensure that the plasmon resonance peak energy is equal to the exciton resonance energy. At $t_f < 1.0$ nm, one peak is observed near the original plasmon resonance. After increasing $t_f$ above 5.0 nm, the $\sigma_{sca}(\omega)$ spectra broaden, and two peaks appear at the edges of the exciton resonance spectra. This spectral broadening indicates the growing of two types of coupled modes between a plasmon and an exciton: an in-phase coupled mode and an out-of-phase coupled mode, which correspond to the constrictive and destructive interferences between a plasmon oscillator and an exciton oscillator, respectively.[8,13] As $t_\text{f}$ increases, the energy gap between the two coupled modes widens due to the increase in the oscillator strength of the exciton of the dye film. The coupled modes are manifested in the form of two peaks when the energy gap exceeds the spectral width of the original exciton resonance of the dye film. This phenomenon suggests that the coupling energy between plasmons and excitons becomes larger than their dephasing energies of around 100 and 500 meV, respectively, indicating a switch from weak coupling to strong coupling. The split energy between the two peaks is proportional to the coupling energy between plasmons and excitons.[8,13,14] The asymmetries of the spectral shapes of coupled resonance are induced by the asymmetric shape of the exciton resonance spectrum of the dye films reflecting their multi-level electronic states.[9]

Figure 3(a2) shows the changes in the $\sigma_\text{sca}(\omega)$ spectra of the coated dimer induced by increasing $t_\text{f}$ from 0 to 0.5 nm. The value of $d$ is set to 1.2 nm to make the plasmon resonance peak energy equal to the exciton resonance energy. The two peaks are clearly observed at the edges of the exciton resonance spectrum even at a film thickness of 0.2 nm. Because the volume of the R6G coating is approximately equal to that of ellipsoidal NPs at the same film thickness, the appearance of these two peaks indicates that the EM coupling per film volume between a plasmon and an exciton for the coated dimers is around 25 (5.0 nm / 0.2 nm) times stronger than that for the coated ellipsoidal NPs. According to the cavity QED, this phenomenon can be explained by the high amplitude of the vacuum field $E_{vac} = \sqrt{\dfrac{\hbar\omega}{2\varepsilon_0\varepsilon_m V}}$ confined inside the gap, where the mode volume $V$ is much smaller than its value for ellipsoidal NPs.[4] The unclear high-energy peaks as compared with those of ellipsoidal NPs may be due to the multiple EM coupling between the dipolar plasmon, higher order plasmons, and dye excitons with their phonon replica.

By the way, the factor of 25 obtained by comparing the $t_f$ values required for strong coupling is likely an underestimation because only the molecules localized in the dimer gap interact with the dipole plasmon.[6,28,29] Meanwhile, the molecules located on the large surface of ellipsoidal NPs can interact with the dipole plasmon.[7,22] Thus, a realistic number of molecules interacting with the dipole plasmon should be determined for dimers. We can approximately estimate the number of molecules involved in strong coupling at $t_f = 0.2$ nm. The results of several numerical calculations indicate that the EM field is confined within the NP surface area of 2.0×2.0 nm² of NP ($r$ ~ 30 nm) surface areas at a gap with $d$ of ~ 1.0 nm.[6,28,29] Thus, the volume of the film is equal to approximately ~2 × 0.2 × 4.0 × 10$^{-28}$ m³. At a molar mass of 479 g/mol (~8×10$^{-22}$ g per molecule) and R6G density of $1.26\times10^6$ g/m$^3$, the calculated number of molecules inside the gap is below 3.0, which is consistent with or rather overestimation for both SM strong coupling and SM SERRS at hotspots.[15,30,31] We consider that the sub-nanometer atomic structures inside the gap confines the field, resulting in both SM strong coupling and SM SERRS.[32]

Figure 3(b) displays the $t_f$ dependences of the coupled resonance peak energy determined for ellipsoidal NPs and the dimers. It shows that the low-energy peaks obtained for the dimers are much more sensitive to $t_f$ than the peaks obtained for ellipsoidal NPs. In contrast, the high-energy peaks recorded for both ellipsoidal NPs and the dimers are insensitive to $t_f$. The asymmetric behaviors of the peaks indicate that the high energy of exciton resonance (> 3.0 eV) prevents the blue shifts of the high-energy peaks, owing to the repulsion between the high-energy coupled resonance and high-energy exciton resonance. In other words, the multi-level properties of the dye molecular exciton resonance induce the asymmetric spectral behavior of the coupled resonances in the strong coupling region. In the case of the dimers, this complexity is enhanced by the multiple coupling between the dipolar plasmon, higher order plasmons, and multiple dye excitons including their replica at 2.5 and > 3.0 eV. The complexity of the spectral changes induced by the multi-level properties of molecules is also related to the strong coupling between the surface plasmon and the dye films.[33]

Figure 3(c) shows the peak energies of the coupled resonances plotted against $\sigma_{sca}(\omega)$ for the dimers and ellipsoidal NPs. Before the appearance of the two peaks, the cross-sections decrease with increasing $t_f$, that corresponds to the absorption by the dye films $\text{Im}[N_{R6G}]$. After the peak appearance, the rate of decrease is reduced because the peaks become separated from the absorption peak region of the dye films. The peak intensities of $\sigma_{sca}(\omega)$ obtained for the dimers are more than ten times smaller than those of ellipsoidal NPs because the length of the dipole confined inside the gap is much smaller than the length of the dipole of ellipsoidal NPs. This confinement is the main reason for the dipole localization in the dimer gap, which is more sensitive to the exciton of the dye films than to the dipole of ellipsoidal NPs.[6,28,29]

To investigate the anti-crossing properties, the plasmon resonance of ellipsoidal NPs is tuned against the exciton resonance by varying their aspect ratio. Figures 4(a) and 4(b) show the changes in $\sigma_{sca}(\omega)$ observed with increasing aspect ratio from 0.13 to 1.0. At $t_f$ = 0.5 nm (Fig. 4(a)), the peaks just decrease their intensities due to the film absorption during the crossing of the exciton resonance at around 2.4 eV, indicating that the system is in a weak coupling region. At $t_f$ = 10 nm, the peaks cannot cross the exciton resonance, as shown in Fig. 4(b). Instead of crossing, they jump over the exciton resonance and start to grow again from its edge of the exciton resonance, suggesting that the system is in a strong coupling region. To obtain more insights, the relationships between the peak energies, peak intensities, and aspect ratios are investigated. Figure 4(c) shows the relationships between the peak energies and aspect ratios, while Fig. 4(d) displays the relationships between the peak energies and peak intensities. At $t_f < 4$ nm, the peaks can cross the exciton resonance with redshifting by < 80 meV from the original positions and intensity reduction. At $t_f > 5$ nm, the peaks redshift by > 80 meV, and the two peaks appear close to the exciton resonance region with decreasing intensities. As $t_f$ increases, the peaks clearly repulse from the exciton resonance spectrum, which is characteristic of the vacuum Rabi splitting (VRS) caused by the strong coupling between a plasmon and an exciton. In summary, Figs. 4(c) and 4(d) clearly show that weak coupling is switched to strong coupling with increasing $t_f$ in the elastic scattering spectra.

The anti-crossing behavior of the dimers at different gap distances is examined as well. Figures 5(a) and 5(b) show the changes in $\sigma_{sca}(\omega)$ observed for the coated dimers while increasing $d$ from 1.2 to 12.0 nm. At $t_f = 0.05$ nm (Fig. 5(a)), the peaks can cross the exciton resonance with decreasing their intensities, indicating that the system is in a weak coupling region. At $t_f = 0.5$ nm (Fig. 5(b)), the two peaks appear near the edges of the exciton resonance instead of crossing, suggesting that the system is in a strong coupling region. Unlike the case of ellipsoidal NPs, the higher energy peak looks unclear due to the disturbance of higher order plasmons. Furthermore, the relationships between the peak energies, peak intensities, and gap distances are obtained as well. Figure 5(c) shows the relationships between the peak energies and gap distances, while Fig. 5(d) displays the relationship between the peak energies and peak intensities. They show that the behaviors of the dimer peaks significantly differ from those of ellipsoidal NPs. At $t_f < 0.2$ nm, the peaks first exhibit redshifts by 250 meV and then return to their original positions with higher intensities as $d$ increases (their intensities are reduced during the crossing of the exciton resonance). These phenomena indicate that the system in the weak coupling region becomes more weakly coupled with increasing $d$. At $t_f > 0.3$ nm, the two peaks appear, and the low-energy peaks redshift by $> 270$ meV. They also jump over the exciton resonance during crossing and start to grow again from its edge with higher intensities. Despite the appearance of the two peaks corresponding to strong coupling, they do not exhibit distinct repulsing behavior, which suggests that strong coupling is switched to weak coupling with increasing $d$. In summary, Figs. 5(c) and 5(d) indicate that the switch from weak coupling to strong coupling at higher $d$ makes the anti-crossing process less visible in spectral changes.

To resolve the unclearness of the anti-crossing properties at different gap distances, plasmon resonance should be tuned without changing the morphology of the dimers. In this work, the anti-crossing behavior of the coated ellipsoidal NPs is examined while changing the refractive indexes of the surrounding media. Figures 6(a) and 6(b) show the changes in $\sigma_{sca}(\omega)$ observed at $t_f$ values of 0.5 and 10 nm respectively, while decreasing $N_m$ from 2.0 to 1.0. The resulting changes are almost identical to those depicted in Figs. 4(a) and 4(b), which exhibit crossing (weak coupling) and anti-crossing (strong coupling), respectively. The relationships between peak energies, peak intensities, and $N_m$ displayed in Figs. 6(c) and 6(d) are almost identical to those in Figs. 4(c) and 4(d) indicating that the repulsion of the two peaks from the exciton resonance in Fig. 6(c) is due to VRS. In summary, Figs. 6(c) and 6(d) clearly show the switch from weak coupling (crossing) to strong coupling (anti-crossing) in the film thickness dependence of the elastic scattering spectra during plasmon resonance tuning with respect to $N_m$.

The anti-crossing properties of the dimers at various values of $N_m$ are studied as well. Figures 7(a) and 7(b) show the changes in the elastic scattering spectra of the coated dimers observed with decreasing $N_m$ from 2.0 to 1.0. Here, we focus on the dipole plasmon resonance occurred in $\sigma_{sca}(\omega)$ at the lowest energy region. At $t_f$ = 0.05 nm (Fig.

7(a)), the peak centered at 1.6 eV blueshifts with intensity reduction during crossing the exciton resonance, indicating that the system is in the weak coupling region. At $t_f = 0.5$ nm (Fig. 7(b)), the peak at 1.6 eV blueshifts with jumping over the exciton resonance and starts to grow again from its edge, suggesting that the system is in the strong coupling region. The observed properties of the dipole plasmon resonance are similar to those depicted in Figs. 6(a) and 6(b). However, the higher order plasmon resonance makes the behavior of the dipole plasmon resonance very complex. Thus, to obtain insights from the observed spectral changes, the relationships between the peak energies, peak intensities, and $N_m$ are investigated. Figure 7(c) shows the relationships between the peak energies and $N_m$, while Fig. 7(d) displays the relationships between the peak energies and peak intensities. At $t_f < 0.2$ nm, the peaks can cross the exciton resonance with redshifting by < 80 meV from the original positions and reduction of their intensities. At $t_f > 0.3$ nm, the peaks redshift by > 150 meV, and the two peaks appear close to the exciton resonance region with decreasing their intensities. As $t_f$ increases, the peaks clearly repulse from the exciton resonance spectrum, which is a VRS characteristic. In summary, Figs. 7(c) and (d) clearly show the switch from weak to strong coupling with increasing $t_f$ in the spectral changes during plasmon resonance tuning with respect to $N_m$.

## IV. SUMMARY

By combining classical electromagnetism with experimentally obtained parameters, the anti-crossing characteristics of the elastic scattering spectra recorded for the dye-coated silver ellipsoidal NPs and NP dimers are examined while changing the thicknesses of the dye films, their morphologies, and the refractive index of the surrounding medium. The analysis of the observed spectral changes reveals that the dipole plasmon localized in the gap is coupled with the dye exciton of the film 25 times stronger than the dipolar plasmon of ellipsoidal NPs at a gap distance of around 1 nm. The anti-crossing properties of the dimers become more complex due to the multiple strong coupling of the dipole and higher order plasmons localized inside the gap with the multiple molecular excitons of the coated films. The most effective way to observe the anti-crossing behavior in coupled resonance of the dimers is performing plasmon resonance tuning by changing the refractive index of the surrounding medium.

In the calculations conducted using the principles of classical electromagnetism, the dye molecules are represented by a thin film indicating that the dye molecules are assumed to be the collective system under the continuum approximation. However, the realistic number of dye molecules in the dimer gap indirectly estimated from the physical properties of R6G is around three. This phenomenon indicates that near SM SERRS hotspots are potential candidates for exploring near SM strong coupling and the related unique photophysical and photochemical processes including population pumping and ultrafast fluorescence.[16,17,34,35,36]

## ACKNOWLEDGMENTS

This work was supported by the JSPS KAKENHI Grant-in-Aid for Scientific Research (C), number 18K04988.

**Figure captions**

Fig. 1 (a1) Experimental plasmon resonance spectra obtained before (red curve) and after (black curve) the loss of SERRS activity. (a2) Calculated plasmon resonance spectra with coupling energies of 500 meV (red curve) and 0 meV (black curve). (b) A scanning electron microscopy image of a typical silver NP dimer exhibiting SERRS with a magnified image of R6G molecules. (c) Strong coupling between the plasmon and molecular resonances. |g> and |e> are the ground and excited states of the two-level system, respectively. |0> and |1> are the zero-photon and one-photon states of the plasmon resonator, respectively. $2\hbar g$ indicates the energy split of the excited state of the strong coupling system formed by the two-level system and plasmonic resonator. The red and blue arrows indicate the two coupled resonance energies. (d) Calculated anti-crossing properties of the strong coupling between the plasmon and exciton with a coupling energy of 400 meV. Inset: absorption spectrum of the aqueous solution of R6G (green curve).

Fig. 2 (a1) Silver NPs adsorbed by the dye molecules (left) and coated with the dye film (right). The red spots in the left image represent R6G molecules. (a2) A prolate ellipsoidal silver NP with the short and long radii equal to $a$ and $b$, respectively. (a3) A dimer of spherical silver NPs with radius $r$ and gap distance $d$. (b1) and (b2) Complex refractive indexes of silver ($N_{Ag}$) and solid R6G ($N_{R6G}$), respectively. The blue and red spectra represent the real and imaginary parts, respectively. (c1) and (c2) $\sigma_{sca}(\omega)$ values of the uncoated ellipsoidal NP (aspect ratio: 0.42) and NP dimer ($d = 1.2$ nm).

(d1) Aspect ratio dependences of $\sigma_{sca}(\omega)$ obtained for ellipsoidal NPs (black lines) with Im[$N_{R6G}$] (green curve). (d2) $d$ dependences of $\sigma_{sca}(\omega)$ obtained for the dimers (black lines) with Im[$N_{R6G}$] (green curve). (e1) $N_m$ dependences of $\sigma_{sca}(\omega)$ obtained for ellipsoidal NPs (black lines) with Im[$N_{R6G}$] (green curve). (e2) $N_m$ dependences of $\sigma_{sca}(\omega)$ obtained for the dimers (black lines) with Im[$N_{R6G}$] (green curve).

FIG. 3 (a1) Changes in the spectra of $\sigma_{sca}(\omega)$ recorded for the coated ellipsoidal NPs in the $t_f$ range from 0.0 to 10.0 nm (10, 5, 3, 2, and 1 nm: red curves; 0.5, 0.4, 0.3, 0.2, and 0.1 nm: orange curves; and 0 nm: black curve) with Im[$N_{R6G}$] (green curve). (a2) Changes in the spectra of $\sigma_{sca}(\omega)$ recorded for the coated dimer in the $t_f$ range from 0 to 0.5 nm (0.5, 0.4, 0.3, 0.2, and 0.1 nm: orange curves and 0 nm: black curve) with Im[$N_{R6G}$] (green curve). (b) $t_f$ dependences of the coupled resonance peak energy obtained for the dimers (orange curves with open triangles) and ellipsoidal NPs (red curves with open circles) with Im[$N_{R6G}$] (green curve). (c) Coupled resonance peak energies plotted against the peak values of $\sigma_{sca}(\omega)$ for the dimers (orange curves with open triangles) and ellipsoidal NPs (red curves with open circles) with Im[$N_{R6G}$] (green curve). The increases in $t_f$ are indicated by the black arrows.

FIG. 4 (a) and (b) Changes in the spectra of $\sigma_{sca}(\omega)$ recorded for the coated ellipsoidal NPs at $t_f$ values of 0.5 and 10.0 nm and aspect ratios of 0.13 (top panel), 0.42 (middle panel), and 1.0 (bottom panel) with Im[$N_{R6G}$] (green curve). The black curves indicate the spectra obtained for the uncoated ellipsoidal NPs. (c) $t_f$ dependences of the peak energies of $\sigma_{sca}(\omega)$ plotted against various aspect ratios ranging from 0.13 to 1.0 with Im[$N_{R6G}$] (green curve). The red symbols with lines ○, □, ∆, ∇, ◊ correspond to $t_f$ magnitudes of 10, 8, 5, 3, and 2 nm, respectively. The black symbol with line ○ corresponds to $t_f$ = 0 nm. Inset: the red symbols with lines ◊ and + correspond to $t_f$ values of 2 and 1 nm, respectively. The orange symbols with lines ○, □, ∆, ∇, and ◊ correspond to $t_f$ magnitudes of 0.5, 0.4, 0.3, 0.2, and 0.1 nm, respectively. (d) $t_f$ dependences of the peak energies plotted against peak intensities at various aspect ratios from 0.13 to 1.0 with Im[$N_{R6G}$] (green curve). The red symbols with lines ○, □, ∆, ∇, ◊, and + correspond to $t_f$ values of 10, 8, 5, 3, 2, and 1 nm, respectively. The orange symbols with lines ○, □, ∆, ∇, and ◊ represent $t_f$ magnitudes of 0.5, 0.4, 0.3, 0.2, and 0.1 nm, respectively. The black symbol with line ○ corresponds to $t_f$ = 0 nm.

FIG. 5 (a) and (b) Changes in the spectra of $\sigma_{sca}(\omega)$ recorded for the coated dimer at $t_f$ values of 0.05 and 0.5 nm and $d$ magnitudes of 1.2 nm (top panel), 2.8 nm (middle panel), and 12 nm (bottom panel) with Im[$N_{R6G}$] (green curve). The black curves indicate the spectra obtained for the uncoated dimer. (c) $t_f$ dependences of the peak energies of $\sigma_{sca}(\omega)$ plotted at various values of $d$ ranging from 1.2 to 12 nm with Im[$N_{R6G}$] (green curve). The orange symbols with lines ○, □, Δ, ▽, and ◊ represent $t_f$ values of 0.5, 0.4, 0.3, 0.2, and 0.1 nm, respectively. The black symbol with line ○ corresponds to $t_f$ = 0 nm. (d) $t_f$ dependences of the peak energies plotted against peak intensities at various values of $d$ ranging from 1.2 to 12 nm with Im[$N_{R6G}$] (green curve). The orange symbols with lines ○, □, Δ, ▽, and ◊ correspond to $t_f$ values of 0.5, 0.4, 0.3, 0.2, and 0.1 nm, respectively. The black symbol with line ○ corresponds to $t_f$ = 0 nm.

FIG. 6 (a) and (b) Changes in the spectra of $\sigma_{sca}(\omega)$ recorded for the coated ellipsoidal NPs at $t_f$ values of 0.5 and 10.0 nm and $N_m$ magnitudes of 2.0 (top panel), 1.4 (middle panel), and 1.0 (bottom panel) with Im[$N_{R6G}$] (green curve). The black curves indicate the spectra obtained for the uncoated ellipsoidal NPs. (c) $t_f$ dependences of the peak energies of $\sigma_{sca}(\omega)$ plotted against $N_m$ in the range from 2.0 to 1.0. The red symbols with lines ○, □, Δ, ▽, ◊ correspond to $t_f$ values of 10, 7, 5, 3, and 1 nm, respectively. The black symbol with line ○ corresponds to $t_f$ = 0 nm. Inset: the red symbols with lines ◊ correspond to $t_f$ = 1 nm. The orange symbols with lines ○, □, Δ, ▽, and ◊ denote $t_f$ values of 0.5, 0.4, 0.3, 0.2, and 0.1 nm, respectively. (d) $t_f$ dependences of the peak energies plotted against peak intensities in the $N_m$ range from 2.0 to 1.0 with Im[$N_{R6G}$] (green curve). The red symbols with lines ○, □, Δ, ▽, and ◊ correspond to $t_f$ values of 10, 7, 5, 3, and 1 nm, respectively. The black symbol with line ○ corresponds to $t_f = 0$ nm. The orange symbols with lines ○, □, Δ, ▽, and ◊ correspond to $t_f$ values of 0.5, 0.4, 0.3, 0.2, and 0.1 nm, respectively.

FIG. 7 (a) and (b) Changes in the spectra of $\sigma_{sca}(\omega)$ (red lines) recorded for the coated dimers at $t_f$ values of 0.05 and 0.5 nm and $N_m$ magnitudes of 2.0 (top panel), 1.4 (middle panel), and 1.0 (bottom panel) with Im[$N_{R6G}$] (green curve). The black curves indicate the spectra obtained for the uncoated dimers. (c) $t_f$ dependences of the peak energies of $\sigma_{sca}(\omega)$ plotted against aspect ratios in the $N_m$ range from 2.0 to 1.0 with Im[$N_{R6G}$] (green curve). The orange symbols with lines ○, □, Δ, ▽, and ◊ correspond to $t_f$ values of 0.5, 0.4, 0.3, 0.2, and 0.1 nm, respectively. The black symbol with line ○ corresponds to $t_f = 0$ nm. (d) $t_f$ dependences of the peak energies plotted against peak intensities in the $N_m$ range from 2.0 to 1.0 with Im[$N_{R6G}$] (green curve). The orange symbols with lines ○, □, Δ, ▽, and ◊ correspond to $t_f$ values of 0.5, 0.4, 0.3, 0.2, and 0.1 nm, respectively. The black symbol with line ○ corresponds to $t_f = 0$ nm.

Fig. 1

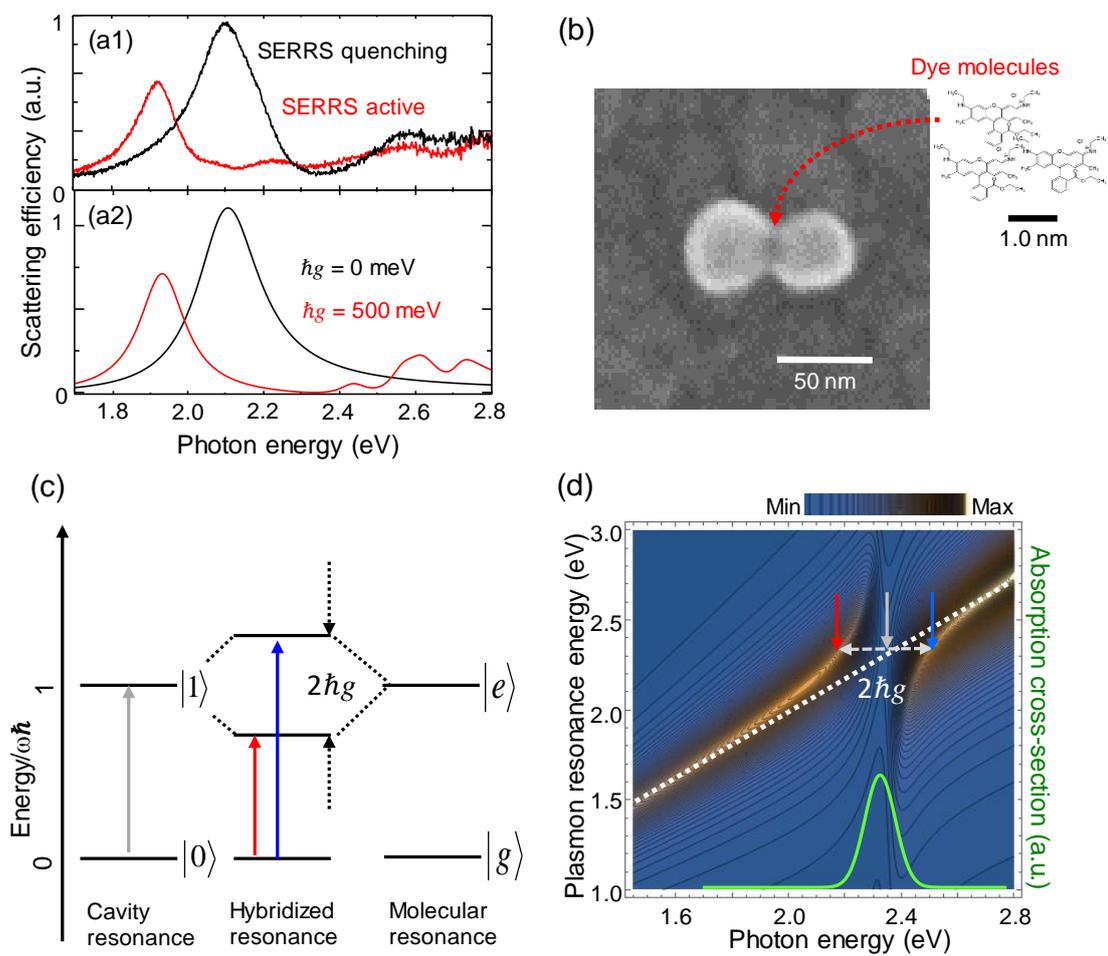

Fig. 2

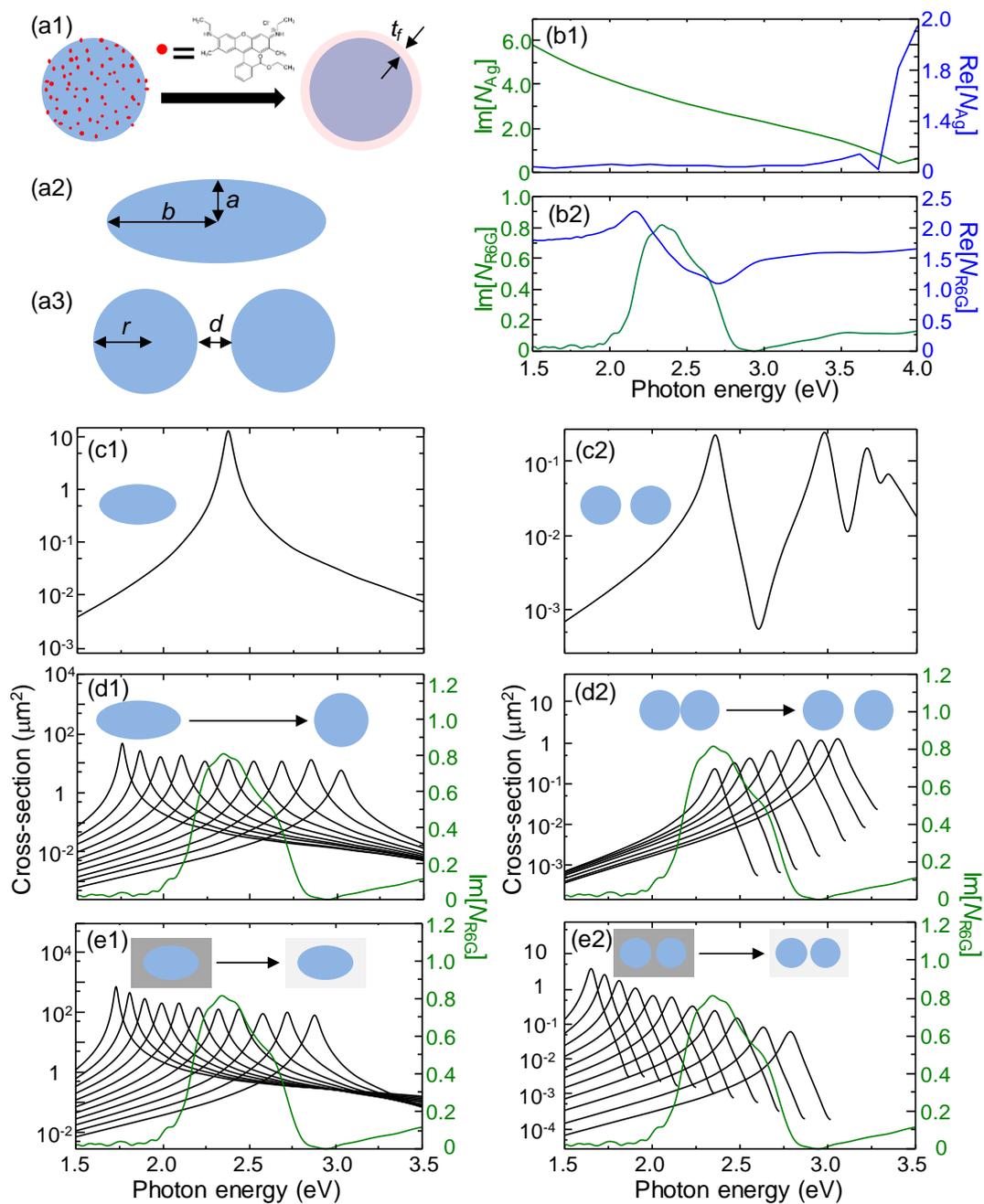

Fig. 3

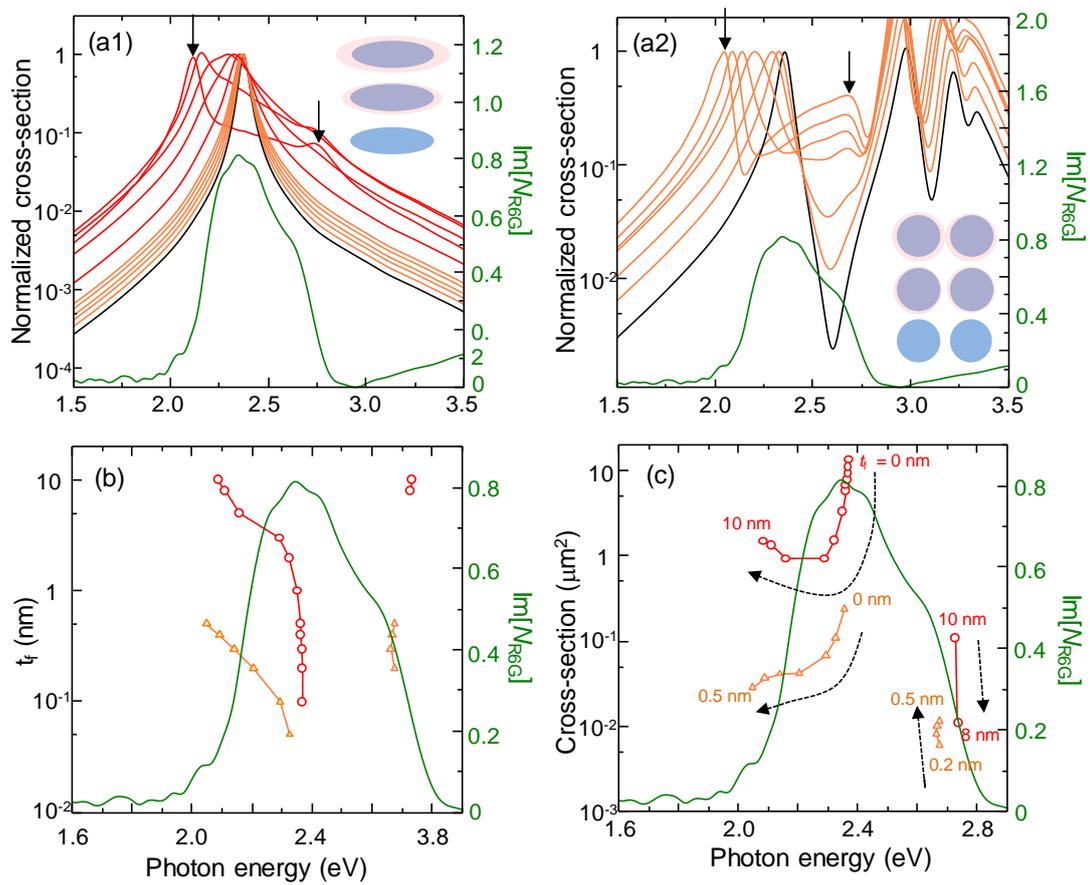

Fig. 4

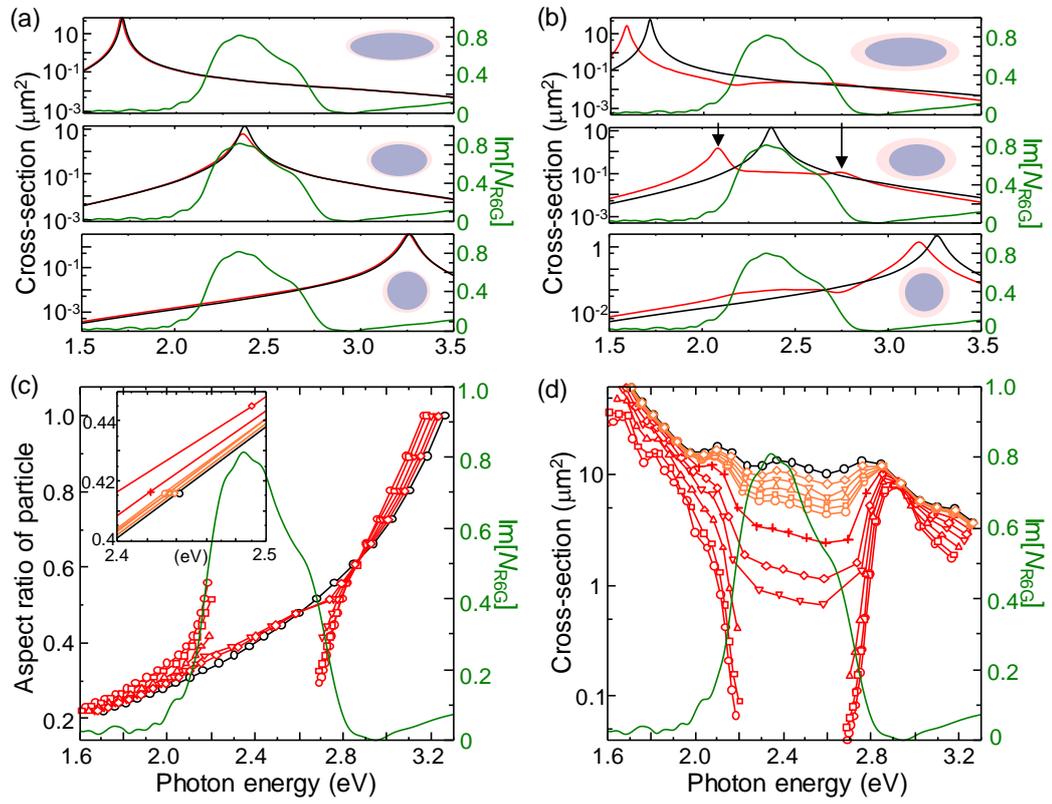

Fig. 5

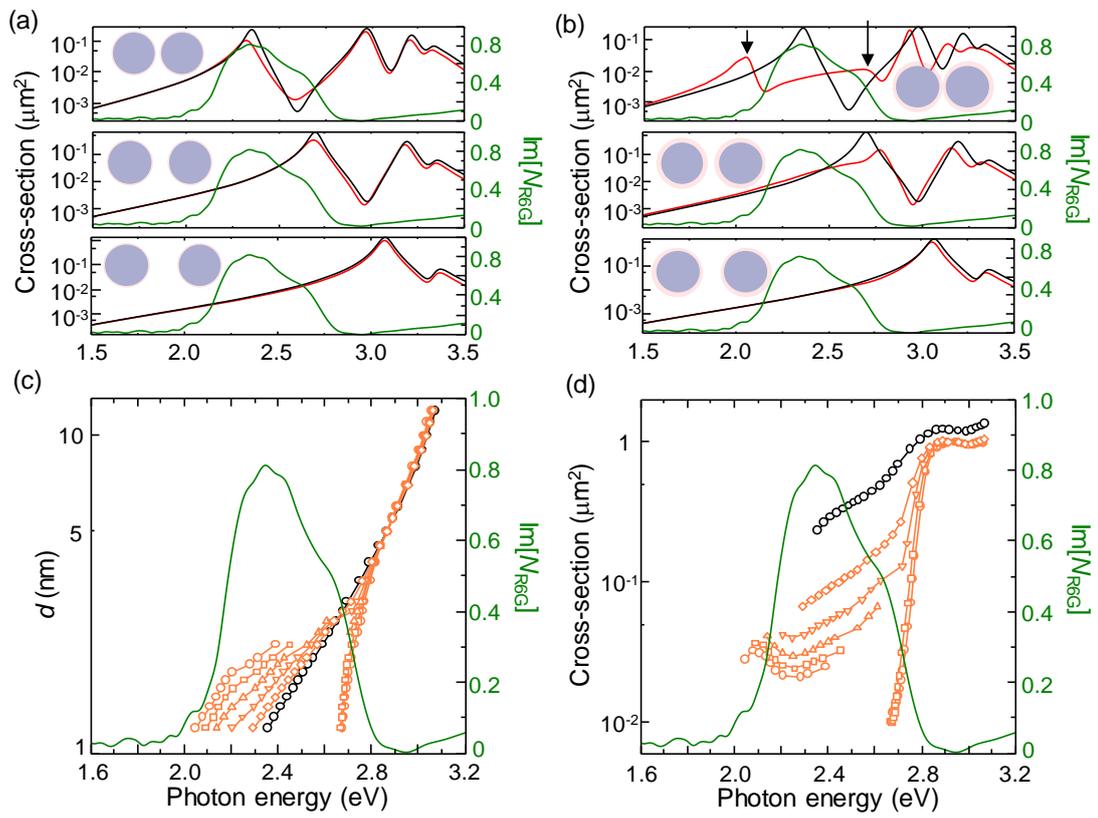

Fig. 6

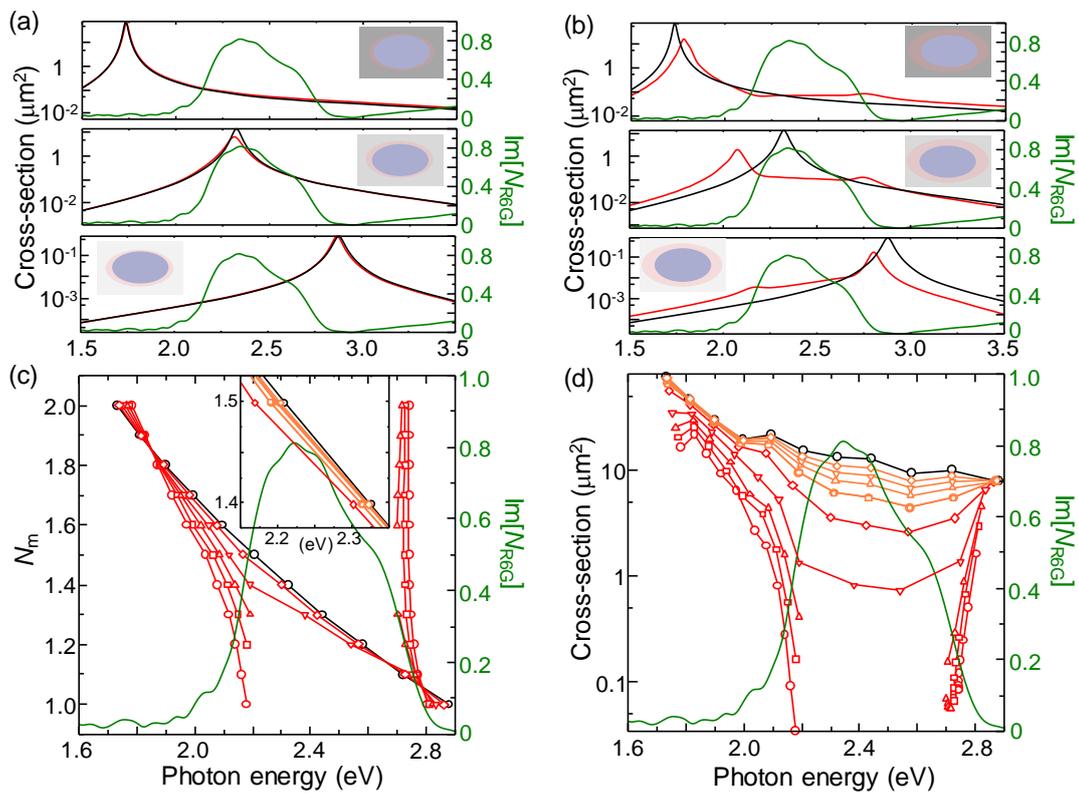

Fig. 7

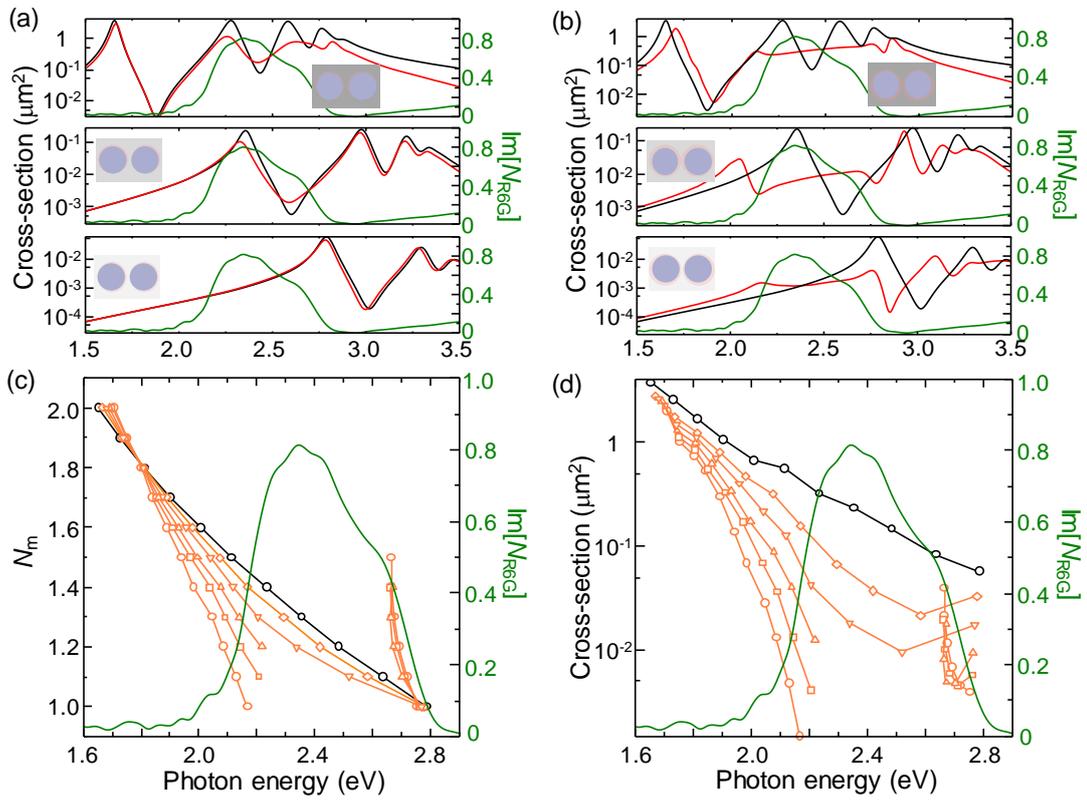